# ARTIFACT REDUCTION IN MULTICHANNEL PERVASIVE EEG USING HYBRID WPT-ICA AND WPT-EMD SIGNAL DECOMPOSITION TECHNIQUES

*Valentina Bono, Wasifa Jamal, Saptarshi Das, and Koushik Maharatna*

School of Electronics and Computer Science, University of Southampton, Southampton SO17 1BJ, UK
`{vb2a12,wj4g08,sd2a11,km3}@ecs.soton.ac.uk`

## ABSTRACT

In order to reduce the muscle artifacts in multi-channel pervasive Electroencephalogram (EEG) signals, we here propose and compare two hybrid algorithms by combining the concept of wavelet packet transform (WPT), empirical mode decomposition (EMD) and Independent Component Analysis (ICA). The signal cleaning performances of WPT-EMD and WPT-ICA algorithms have been compared using a signal-to-noise ratio (SNR)-like criterion for artifacts. The algorithms have been tested on multiple trials of four different artifact cases *viz.* eye-blinking and muscle artifacts including left and right hand movement and head-shaking.

*Index Terms*—Artifact reduction, EMD, ICA, muscle artifact, pervasive EEG, wavelet packet transform

## 1. INTRODUCTION

Electroencephalogram (EEG) is a cost-effective and non-invasive means for monitoring brain signals allowing assessment of cognitive functionalities of a subject [1-2]. However its major problem is its susceptibility of getting corrupted by unwanted artifacts like eye-blinks, eye-ball movement, respiration and cardiac activity at the time of recording rendering the acquired data unusable. To avoid this, typically EEG is captured with the subjects sitting in a constrained position performing cognitive tasks with minimal body movement. This restricts the number of possible cognitive tasks that can be used for effectively studying the brainwave behavior. In addition, when used for studying neurobiological disorders, e.g. autism spectrum disorder (ASD), the presence of the EEG acquisition system in constrained clinical environment may modulate the patient's original brainwaves for the evoked potential due to the patient's subconscious awareness of such system. An unobtrusive EEG system, allowing subject monitoring in a more natural environment during execution of a wide range of cognitive tasks, may allow the clinicians gain better understanding of the subject's cognitive states. However data acquired by such a system is more prone to be corrupted with much severe artifacts compared to the traditional EEG system. The more degrees of freedom allowed for movement introduce a wide variety of muscular artifacts (e.g. head-shaking, jaw movement, hand/body movement etc.) that are not typical in traditional EEG systems. Therefore in order to make the data from such a pervasive system [3] usable in practice, artifact suppression or separation is key.

Although a wealth of research exists exploring possible methods of artifact separation from conventional EEG systems, most of them assume either *a-priori* knowledge of the source of artifacts or use simulated artifacts which in general tend to differ quite remarkably in morphology from the experimentally observed data. In a pervasive EEG system, *a-priori* assumptions are difficult to apply because of potentially enormous number of possible combinations of artifacts resulting from naturalistic movements. Therefore artifact separation for pervasive EEG needs a fundamentally different approach.

In this paper, as a first step of suppressing artifacts in pervasive EEG during natural body movement, we formulate a generalized framework for automatic rejection of eye-blinking, right or left hand movement and head-shaking using two novel hybrid approaches: a) fusion of WPT and ICA (WPT-ICA) and b) fusion of WPT and EMD (WPT-EMD). A SNR-like criterion is introduced as a metric quantifying the signal cleansing performance. Real-life data was captured using Enobio pervasive EEG system [4] over three trials of each artifact mentioned above. Our analysis shows that both of the hybrid algorithms perform better than applying one single technique. In addition, by comparing the SNR-like criterion, the WPT-EMD algorithm gives significantly better results than the WPT-ICA technique in all the cases.

## 2. APPLICABILITY OF TRADITIONAL ARTIFACT SEPARATION IN PERVASIVE EEG

Theoretical studies suggest that an artifact can be removed from a corrupted signal by using linear filtering and linear regression [5] if the source of the artifact can be measured, e.g. eye-blinking artifact removal using Electro-oculogram (EOG) recording, muscular artifact removal using Electromyogram (EMG) recording across specific muscle generating the artifact and using Electrocardiogram (ECG) recoding for the artifact due to heart-beat [1-2, 6-7]. In a pervasive EEG scenario, there is no known way to record the unwanted 'source' signal to pose the artifact removal problem in linear filtering framework. This limits the applicability of using linear combinations of the unwanted artifact signal via adaptive filtering techniques like Least-Mean-Square (LMS) and Wiener filter in Finite/Infinite Impulse Response (FIR/IIR) structure [1, 8]. Among other approaches of artifact suppression, ensemble and robust averaging, adaptive filtering (LMS/Wiener) and Bayesian filtering (including Kalman and particle filters) are quite popular if some *a-priori* knowledge of the statistics or model of the artifact is assumed [1, 8] – a situation less realistic in pervasive EEG system. When the source of the artifact is unknown, blind-source separation (BSS) techniques, researched widely over last few decades, can be applied e.g. ICA [9-11], principal component analysis (PCA) [12] and canonical correlation analysis (CCA) [13-15]. However, in the pervasive EEG system, where there are greater degrees of freedom of movement, a large variety of unknown artifacts may exist [16]. This makes ICA less reliable for separating the artifact component from the EEG signals since the artifacts may come from distributed or multiple sources. Contemporary researchers have also introduced several hybrid techniques to tackle EEG eye-blinking and simulated artifacts e.g. wavelet enhanced ICA [17-20], EMD-ICA [21], ensemble EMD (EEMD)-ICA [22], EEMD-CCA [23], deterministic (wavelet-EMD) and stochastic (ICA-CCA) approaches [24]. But the major limitations of all the above literatures are that the simulations are reported with no experimental artifact but with some well-behaved theoretically tractable simulated artifacts. Our proposed methods perform the artifact separation in a holistic way and irrespective of the type of artifact which affects the signal.

## 3. THEORETICAL FRAMEOWRK FOR THE HYBRID WPT-ICA AND WPT-EMD ALGORITHM

### 3.1. Wavelet Packet Transform (WPT)
WPT is a generalized form of Discrete Wavelet Transform (DWT). DWT decomposes a signal by passing it through a series of high and low-pass filters. This effectively allows the analysis of the signal at different scales and time resolutions. During wavelet decomposition, the high-pass and the low-pass wavelet coefficients are subsampled to give the detail *d[n]* and the approximate *a[n]* coefficients respectively. This routine is applied recursively on the approximation until desired decomposition level is reached. On the other hand, in WPT the decomposition is applied in both low and high frequency coefficients, i.e. both the detail and the approximation branches are transformed to get nodes for all the decomposed levels [25].

### 3.2. Independent Component Analysis (ICA)

ICA is a technique that uses the principles of statistical independence to find a representation in which the components are independent. Fast-ICA, which has been applied in the hybrid WPT-ICA algorithm, implements a fixed-point iteration scheme for maximizing the negentropy approximation for non-Gaussianity [26]. Data-centric polynomial contrast functions ( $G$ with first derivative $g$ and second derivative $g'$ ) can be used to estimate the inverse matrix, $W$ (inverse of mixing matrix $A$ ). The iteration step of the Fast-ICA algorithm is given in (1), where $w$ is a single column of $W$, and $z$ is the whitened and centered data.

$$w \leftarrow E\{zg(w^T z)\} - E\{g'(w^T z)\}w \quad (1)$$

ICA can be used to detect which channels are most affected by the artifacts. For example, the eye-blinking artifact corrupts the frontal lobe electrodes as shown in [16].

### 3.3. Empirical Mode Decomposition (EMD)
EMD decomposes a nonlinear and non-stationary signal *x(t)* into a finite number of intrinsic mode functions (IMFs) $h_i(t)$ and a residual *r(t)*, as in (2).

$$x(t) = \sum_{i=1}^{N} h_i(t) + r(t) \quad (2)$$

Each IMF satisfies the following two conditions [27]:
i) In the original signal the number of extrema and the number of zero-crossings must either equal or differ at most by one
ii) At any point the mean value of the envelope defined by the local maxima and local minima is zero

The upper envelope *u(t)* and the lower envelope *l(t)* of the signal *x(t)* is calculated. Then, the mean envelope *m(t)* is calculated as $m(t) = [u(t) + l(t)]/2$ which is then subtracted from original signal, i.e. $h(t) = x(t) - m(t)$. Once the *h(t)* satisfies the conditions mentioned above, it will be subtracted from the signal and the process will be repeated until the residual will be reached. This enables identifying the basic irregular components of the corrupted signal and the artifacts in the EEG signal can be identified as IMFs and hence can be rejected to clean the signal.

### 3.4. Proposed Hybrid Algorithms
Fig. 1 shows the schematic diagram of the two proposed algorithms along with different parameters which are judged to quantify the effect of the artifact and the rules that are applied to automatically reject them. In both of the proposed methods, firstly the WPT is applied and the nodes of the seventh decomposition level are considered for the identification of the one containing the maximum effect of the artifact. Although here the WPT is applied on each EEG channel, it takes the Wavelet coefficients from all the channels into account in order to identify the maximum corrupted node in wavelet packet decomposition. The maximum standard deviation of the wavelet energy $\sigma_E$ across all the 19 channels is considered as the criterion to find out which node has maximally captured the effect of the artifact. The coefficients of that particular node are then rejected while reconstructing the cleaned signal. The obtained WPT-cleaned signal is then analyzed by the ICA (WPT-ICA) and the EMD (WPT-EMD), as shown in Fig 1.

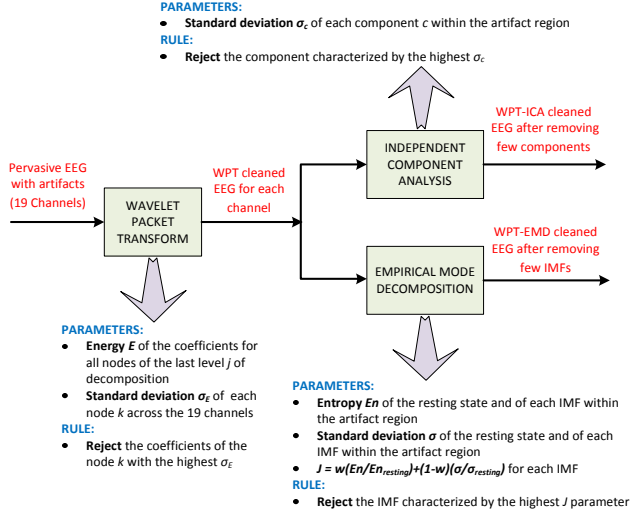

Fig.1. Schematic diagram of the two hybrid algorithms.

In WPT-ICA, the Fast-ICA algorithm is applied on all the 19-channels with the aim of separating the global effect or the common components of the artifact corrupting all the channels. The temporal standard deviation $\sigma_c$ of each component within the artifact region (found by suitable amplitude threshold) is used as a criterion to identify the component containing the artefact. This is characterized by the highest $\sigma_c$ and it is considered as the most influential part of the artifact, affecting some or all of the channels and therefore rejected during the signal reconstruction.

When using WPT-EMD technique, EMD is applied as a second step on the WPT-cleaned signal instead of ICA. The EMD, being a nonlinear signal decomposition technique, tries to match the irregular oscillations and the inconsistently large amplitude which are characteristics of all types of artifacts. The IMF with the highest entropy $En$ and temporal standard deviation $\sigma$ is considered the one which fulfills the above criteria. The En and $\sigma$ of each IMF in the corrupted EEG as well as in the resting state (eye-closed) EEG are used to evaluate a hybrid index $J$ for each channel which can capture large oscillations and non-random almost periodic IMFs as potential components of the artifact. $J$ has been formulated as a weighted sum of normalized $En$ and $\sigma$ with respect to their resting state values as shown in (3).

$$J = w\left(En/En_{resting}\right) + (1-w)\left(\sigma/\sigma_{resting}\right) \qquad (3)$$

The IMF with the highest $J$ parameter is considered as the most responsible part in the artifact region which is then rejected during reconstruction of the cleaned signal.

As a quantitative measurement of the extent of artifact cleansing, the following average SNR-like criterion (4) is evaluated for all the channels using both the algorithms. The resting state is acquired while the person has his/her eyes closed, where the artifacts are absent. The corrupted signal is acquired while the person is doing an artifact-related task during the entire EEG recording. While calculating the SNR, the power of the artifact is obtained from the difference (in power) between the corrupted and the clean signal by averaging over all EEG bands, all electrodes and all trials. For evaluating the resting state power, only one trial was considered. According to the criterion (4), lower values indicate better performances. Since, the numerator and $P_{resting}^{avg}$, $P_{corrupted}^{avg}$ are constants, lower value of $P_{clean}^{avg}$ will increase the denominator, which results in reduced SNR-like criterion, as opposite to the convention of the standard formulation of SNR [28].

$$SNR_{avg} = \frac{P_{resting}^{avg}}{P_{artifact}^{avg}} = \frac{P_{resting}^{avg}}{P_{corrupted}^{avg} - P_{clean}^{avg}}$$

$$= \frac{\frac{1}{m}\sum_{i\in\{\delta,\theta,\alpha,\beta,\gamma\}}\sum_{j=1}^{m}P_{resting_i}^{j}}{\frac{1}{mn}\left(\sum_{i\in\{\delta,\theta,\alpha,\beta,\gamma\}}\sum_{j=1}^{m}\sum_{k=1}^{n}P_{corrupted_i}^{j,k} - \sum_{i\in\{\delta,\theta,\alpha,\beta,\gamma\}}\sum_{j=1}^{m}\sum_{k=1}^{n}P_{clean_i}^{j,k}\right)} \qquad (4)$$

[$m$ = # of electrodes, $n$ = # of trials]

### 4. SIMULATION RESULTS

As it is evident in an example of eye-blinking artifact in Fig. 2, in the time domain the peak-to-peak amplitudes of the EEG signal is reduced for both of the algorithms. The algorithm changes the power spectrum of the signal mostly in low frequency. In frequency domain, the high peak below 2 Hz is significantly reduced, while the rest of the spectra remains unaffected. Closer inspection of the high frequency range shows that $\beta$ and $\gamma$ bands are not affected by the WPT-EMD artifact reduction algorithm, whereas, they are slightly modified and hence less reliable for the second algorithm i.e. WPT-ICA.

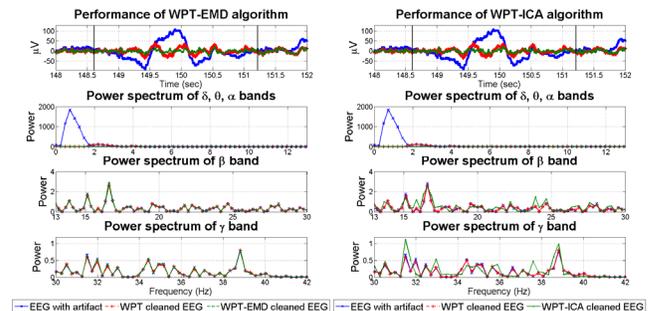

Fig. 2. Example of eye-blinking artifact in channel $FP_1$.

Similar phenomena can be observed for the artifacts due to a right hand movement in Fig. 3 where the time domain oscillations are reduced and clear improvement in the low frequency bands like $\delta$, $\theta$, $\alpha$ can be observed for both of the proposed algorithms. The WPT-ICA still suffers from distorting the high frequency information in a slightly higher extent than the WPT-EMD. For the left hand movement (Fig. 4), in time domain WPT-ICA cannot reduce the large inconsistent jumps or drift in the artifact region, which are decreased to a higher extent using WPT-EMD. It is also evident that the WPT-ICA algorithm essentially increases the power spectrum of the reconstructed signal, especially in $\beta$ and $\gamma$ bands. This is due to the fact that replacing one component with zero may slightly increase the high frequency oscillations if some of the independent components have got opposite signs. This phenomenon can be justified in a way that the independent components are likely to have much wider spectral information than the IMFs which are localized in a narrow frequency range. So, removal of independent components may affect the frequency domain information over a wider frequency range than removal of certain IMFs.

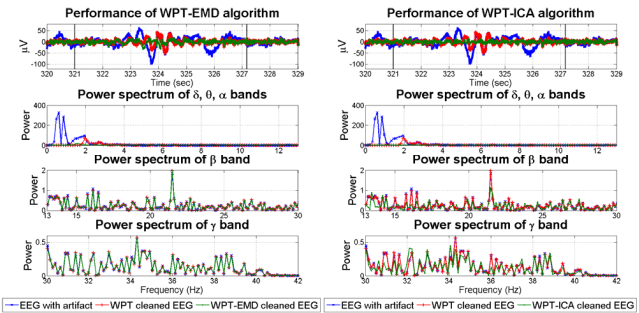

Fig. 3. Example of right hand movement in channel $P_4$.

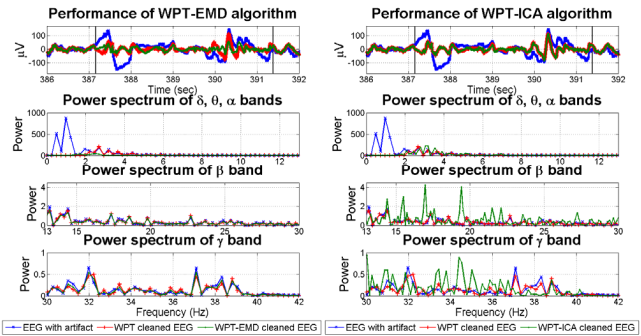

Fig. 4. Example of left hand movement in channel $FP_1$.

As evident from Fig. 5, the head-shaking in a natural scenario will have the most severe effect in terms of corrupting the amplitude of the original EEG signal and it almost pervades all the EEG channels unlike eye-blinking which was restricted only in the frontal electrodes. Removing this particular type of artifact is more challenging than the other cases, because the band of oscillation in time domain is higher compared to the other cases of artifacts.

Also in frequency domain, the peak below 2 Hz, present in all the previous cases, is characterized by very high amplitude. The application of both the proposed algorithms shows good cleaning performance in all the bands except $\delta$ (i.e. below 4 Hz).

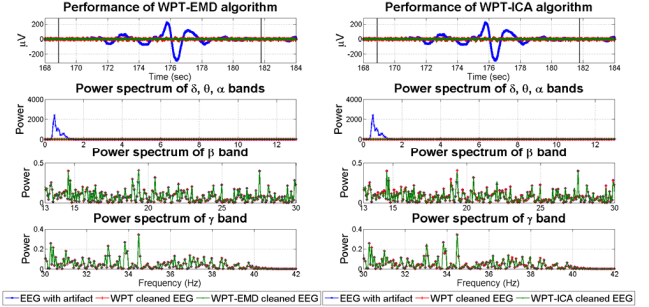

Fig. 5. Example of head shaking artifact in channel $C_4$.

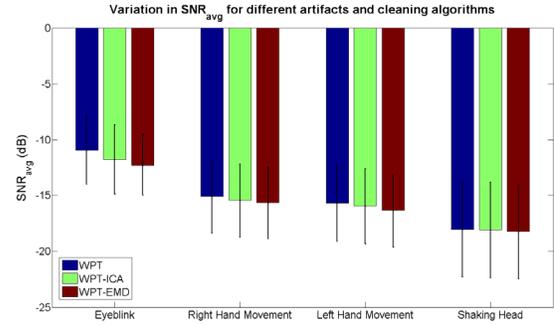

Fig. 6. SNR of different cleaning algorithms for three independent trials of the artifacts.

The SNR values (in dB) have been compared in Fig. 6 for the WPT, WPT-ICA and WPT-EMD cleaned EEG signals respectively while treating the artifact as unwanted noise [28] using the formulation given in (4). Fig. 6 shows that the performance is getting enhanced distinctly from WPT to WPT-ICA and further using WPT-EMD, described by low SNR values for the first three classes of artifacts and improves slightly for the head shaking artifact.

## 5. CONCLUSION

Four types of artifact in the pervasive EEG system have been reduced using two hybrid techniques of signal quality enrichment i.e. WPT-EMD and WPT-ICA. Our approach is robust and holistic since it does not require any *a-priori* information about the artifact and is thus suited for artifact removal in pervasive EEG. Further research can be directed towards establishing the proposed hybrid techniques for other muscular artifacts.

## ACKNOWLEDGMENT

This work was supported by FP7 EU funded MICHELANGELO project, Grant Agreement #288241.